\documentstyle[aps,prb,epsfig]{revtex}
\begin{document}
\author{V.Ya. Demikhovskii \footnote{demi@phys.unn.ru} and
D.V. Khomitskiy}
\address{Nizhny Novgorod State University \\
 Gagarin Ave. 23, Nizhny Novgorod 603950, Russia}
\title{Quantum Hall effect in a {\it p}-type heterojunction with a lateral
surface quantum dot superlattice}

\maketitle
\begin{abstract}
The quantization of Hall conductance in a {\it p}-type heterojunction with
lateral surface quantum dot superlattice is investigated. The topological
properties of the four-component hole wavefunction are studied both in
{\bf r}- and {\bf k}-spaces. New method of calculation of the Hall conductance
in a 2D hole gas described by the Luttinger Hamiltonian and affected by
lateral periodic potential is proposed, based on the investigation of
four-component wavefunction singularities in {\bf k}-space.
The deviations from the quantization rules for Hofstadter "butterfly" for
electrons are found, and the explanation of this effect is proposed.
For the case of strong periodic potential the mixing of magnetic subbands is
taken into account, and the exchange of the Chern numbers between magnetic
subands is discussed.
\end{abstract}

\vspace{0.5cm}
PACS number(s): \ 73.20.Dx, 73.40.Hm, 73.50.-h

\vspace{0.5cm}

\section{Introduction}

Quantum states and transport of 2D Bloch electrons in a magnetic field are
the phenomena which show extremely rich variety of physical and topological
properties. The fascinating physical problems occurring here are caused by
the mutual effects of the lattice periodic potential and the non-periodic
vector potential of a uniform magnetic field. It is known that the former
leads to the energy band structure while the latter tends to form discrete
energy levels. The parameter which plays an important role in the problem is
the magnetic flux $\Phi$ penetrating the lattice elementary cell. If the flux
is equal to a rational number $p/q$ of flux quanta $\Phi_0=2\pi \hbar c/|e|$
where $p$ and $q$ are mutually prime integers, it is possible to define a new
set of translations on the lattice, called magnetic translations \cite{Zak,LP}
for which the quasimomentum is a good quantum number. For example, if the
vector potential of uniform magnetic field $B$ be chosen in Landau gauge
${\bf A}=(0,Bx,0)$, and $\Phi/ \Phi_0=p/q$, the simplest form of magnetic
translations on a square lattice with the period $a$ is
$x\rightarrow x+qna, \em  y\rightarrow y+ma$ where $n$ and $m$ are integers.
So, the magnetic elementary cell is $q$ times larger in $x$ direction, and
the corresponding magnetic Brillouin zone (MBZ) is defined as following:
\begin{eqnarray}
\label{MBZ}
-\pi /qa \le k_x \le \pi /qa, \quad  -\pi /a  \le k_y \le \pi /a.
\end{eqnarray}

\noindent When the quasimomentum runs over the MBZ (\ref{MBZ}), the energy
varies in a band which is called a magnetic subband. When the amplitude of
periodic potential $V_0$ is smaller than the cyclotron energy
$\hbar \omega _c$ one can neglect the influence of neighboring Landau levels
and may obtain the set of $p$ magnetic subbands arising from a single level
\cite{Thouless}. If several electron Landau levels are taken into account,
the periodic potential leads to the mixing between magnetic subbands
originating from different levels \cite{Silb,Geisel,DP}. However, the in-plane
properties of the electron wavefunction remain the same both for coupled and
uncoupled Landau levels. For example, one can see that, regardless to
the particular form and the number of Landau levels taken into account,
the electron wavefunction gains an additional phase under the magnetic
translations. The relation between the translated and the initial
wavefunctions in magnetic field is known as the generalized Bloch conditions
(or Peierls conditions) \cite{Peierls,LP}
\begin{eqnarray}
\nonumber
\psi_{k_x k_y}(x,y,z)=\psi_{k_x k_y}(x+qa,y+a,z)\exp(-ik_x qa)\times \\
\label{pei}
\times \exp(-ik_y a)\exp(-2\pi ipy/a).
\end{eqnarray}

\noindent It follows from (\ref{pei}) that the wavefunction gains the phase
$2\pi p$ after the circulation along the boundary of the magnetic unit cell.
As a result, the magnetic field forces the periodic part
$u_{{\bf k}}({\bf r})=\exp(-i{\bf kr})\psi_{{\bf k}}({\bf r})$ of the
wavefunction to have $-p$  vorticity in the magnetic unit cell which
indicates that there are at least $p$ zeros of the wavefunction per each
magnetic cell \cite{Kohmoto}. This result has a topological nature because of
its independance of the shape and the amplitude of periodic potential.

During last years the researchers have investigated several significant
theoretical and experimental features of the systems where a 2D electron gas
with additional periodic potential is in the regime of quantum Hall effect
(QHE). If a single Landau level is splitted by a 2D periodic potential which
has the area of elementary cell corresponding to $p/q$ of flux quanta
penetrating the cell, the spectrum transforms to the system of $p$ magnetic
subbands grouped near the unperturbed level. One might expect that each of
magnetic subbands gives a Hall conductance $\sigma_H$ equal to $e^2/ph$, but
according to Laughlin each subband must carry an integer multiple of the Hall
current carried by the entire Landau level which is equal to $e^2/h$. For the
first time the confirmation of this rule which describes the quantization of
Hall conductance in periodically modulated 2D systems has been obtained by
Thouless, Kohmoto, Nightingale, and den Nijs in their pioneer paper
\cite{Thouless}. They have studied in detailes a simple quasi-1D model of
a strongly anisotropic lattice for which an explicit expression for $\sigma_H$
has been derived. If the Fermi energy falls into the {\it r}th gap of the
{\it N}th splitted Landau level, the Hall conductance can be written as
following:
\begin{equation}
\nonumber
\sigma_H=\frac{e^2}{h}(t_r+N-1),
\end{equation}

\noindent where $t_r$ is an integer obtained from the Diophantine equation
\begin{equation}
\label{Dioph}
t_rp+s_rq=r.
\end{equation}

\noindent Equation (\ref{Dioph}) has integer solutions for some integer values
of $s_r$, $|s_r|\le p/2$. It was found further that the quantization of
$\sigma_H$ in periodically modulated systems has a topological nature. Namely,
the value of $\sigma_H$ for a fully occupied magnetic subband is related to
the number and the type of the wavefunction singularities in ${\bf k}$-space.
Kohmoto has shown that these singularities determine the first Chern number
for a particular nagnetic subband which is, in units of $-e^2/h$, exactly the
Hall conductance of this subband \cite{Kohmoto}.

An original method for calculation of the Hall conductance of 2D electron
gas affected by weak periodical modulation has been proposed
by Usov \cite{Usov}. He has shown that the value of $\sigma_H$ is related to
the winding numbers $S_m$, $m=1,2,\ldots$ of the wavefunction singularities
in the extended MBZ (\ref{ext}). These singularities are direct
consequence of a non-trivial topology of the MBZ or EMBZ, and a winding number
$S_m$ is determined as the phase mismatch at the beginning and at the end of
a circulation around the singular point ${\bf k}_m$. As a result, the Hall
conductance of a fully occupied subband $\alpha$ is given by: \cite{Usov}
\begin{equation}
\label{usov}
\sigma_H^{\alpha}=\frac{e^2}{h}\left[\frac 1p+\frac qp \sum_m{S_m}\right].
\end{equation}

\noindent The topological features of the problem have been discussed for
the first time by Novikov \cite{Novikov}. Namely, the formation of $p$
magnetic subbands near the Landau level was treated as a fiber bundle of
magnetic Bloch functions on a $T^2$-torus which is the MBZ (\ref{MBZ}). This
problem has also been discussed by Avron, Seiler and Simon using homotopy
theory \cite{ASS}. The generalization of the proof of existance of the
topological invariant to the situation where many-body interaction and
substrate disorder are present has been obtained by Niu, Thouless, and Wu
\cite{NTW}. It should be mentioned that Simon \cite{Simon} made a connection
between the topological invariant and Berry's geometrical phase factor
\cite{Berry} in the quantum adiabatic theorem. The Berry's phase links the
Hall conductance with a 2D integral over the MBZ (\ref{MBZ}) of the so called
Berry curvature which is a ${\bf k}$-dependent function, and it can be written
in the spirit of the Kubo formula for the conductance
\cite{Kohm93,Chang,Kohm02}.

Another approach to the calculation of Hall conductance is based on the
St\v{r}eda formula \cite{Streda}. If the Fermi level is located in the energy
gap, the Hall conductance is given by
\begin{equation}
\label{str}
\sigma_H=ec\frac{\partial N(E)}{\partial B}
\end{equation}

\noindent where $N(E)$ is the number of states per unit area having energy
lower than the gap energy. The formula (\ref{str}) has been widely used for
calculations of the Hall conductance of 2D electron gas with periodical
modulation, even in the presence of Landau level coupling \cite{Geisel}.
It was also applied to 3D systems \cite{Kohm90,Kohm92,Koshino} were
the generalization of (\ref{str}) is known as the Kohmoto-Halperin-Wu formula
\cite{Koshino}. However, the application of St\u{r}eda formula (\ref{str})
to the systems with multi-component wavefunction is not justified and thus
we shall focus on the analytical approach described above.

During last decade the number of experimental studies have been performed in
order to investigate a 2D electron gas laterally modulated by a surface
superlattice of quantum dots (antidots). Such a system is convenient
for investigation of both classical effects (commensurability of the lattice
periods and cyclotron radius, transition to chaos, etc.) and of the energy
spectrum consisting of magnetic subbands. For example, the oscillations of
longitudinal magnetoresistance have been detected under the conditions where
classical cyclotron radius $2R_c$ envelopes an integer number of antidots or
numerous reflections from one antidot occur \cite{Weiss,Eroms}. It should
be mentioned that the effects of randomly distributed impurities on collision
broadening and transport scattering rate in 2D electron gas with periodical
modulation, which consideration is of great importance for experimentators,
have also been studied theoretically \cite{PG}. The first experimental
evidences of electron Landau levels splitted into the set of magnetic subbands
have been obtained by the measuremets of the longitudinal magnetoresistance
\cite{Schl}. Then, the Hall resistance in a laterally modulated 2D electron
gas have been studied experimentally and the confirmations of subband energy
spectrum have been found \cite{Albrecht}.

The experiments in $p$-type heterojunctions without periodic potential have
also become possible due to the progress in technology which substantially
improved the quality of $p$ channels in GaAs/AlGaAs heterojunctions
\cite{Volkov}. Thus, almost all intriguing phenomena found for 2D electron
systems were also observed in 2D hole channels. It should be noted that their
theoretical investigation has been carried out much earlier, for example,
the studies of fractional quantum Hall effect in a 2D hole gas have been
performed \cite{MacDonald}. In several recent publications the
magnetotransport in 2D hole gas with lateral periodic modulation was studied
\cite{Weiss02,DK02}. The corresponding quantum-mechanical model describing the
hole subband spectra and the four-component magnetic Bloch wavefunctions has
been derived by us recently, and the magnetooptical properties of laterally
modulated 2D holes have also been studied \cite{ourPRB}. As soon as the
transport experiments in laterally modulated 2D hole gas have started
\cite{Weiss02}, it is now needful to derive a quantum-mechanical description
of transport phenomena in such systems, and, in particular, of the quantum
Hall effect.

In the current paper we present a new method of calculation of the Hall
conductance in a 2D hole gas affected by lateral periodic potential. In Sec.
II we briefly describe the magnetic hole Bloch states in a $p$ - type
heterojunction subjected to a magnetic field and affected by lateral
periodic potential, which have been studied by us previously \cite{ourPRB}.
In Sec. III we generalize the method derived by Usov \cite{Usov} for
calculation of the Hall conductance in a system studied in Sec. II where the
charged particle is described by a four-component eigenfunction of the
Luttinger Hamiltonian. We find an unusual behavior of the Hall conductance as
a function of the Fermi energy in comparence to the well-known dependance
obtained for Hofstadter "butterfly" \cite{Thouless} for the electrons.
We propose an explanation of this effect by evaluating the role of the
off-diagonal terms of the Luttinger Hamiltonian which provide a highly
non-equdistant character of hole Landau levels. The quantization of the Hall
conductance has been investigated both at weak and at strong periodic
potential. In the latter case we've taken into account the magnetic subband
mixing which leads to the exchange of the Chern numbers between magnetic
subands, changing their impact to the Hall conductance. We believe that the
differences between the quantization of Hall conductance in n- and p-type
heterojunctions which have been predicted by us can be observed
experimentally. In Sec. IV we give the summary of our results.

\section{Magnetic hole states in laterally modulated heterojunction}

The holes are studied near the upper edge of GaAs $p$-like valence band
located at ${\bf k}=0$. We assume that the external magnetic field is
oriented along $\langle 001 \rangle$ crystal direction which is perpendicular
to the heterojunction plane $(xy)$. The 2D holes are described in the
$|J;m_J\rangle$ basis by the $4\times 4$ Luttinger Hamiltonian where both
magnetic field and confinement potential $V_h(z)$ of a single heterojunction
are taken into account \cite{Broido,Volkov}. In addition, the periodic
potential $V(x,y)$ of a quantum dot superlattice is introduced which simplest
form is \cite{DP}
\begin{equation}
\label{vxy}
V(x,y)=V_0\cos^2\frac{\pi x}{a}\cos^2\frac{\pi y}{a}.
\end{equation}

\noindent Here $a$ is the superlattice period and the case $V_0<0\ (>0)$
corresponds to the periodic electric potential generated by quantum dot
(antidot) superlattice. The Hamiltonian for magnetic Bloch hole quantum
states written in atomic units $\hbar=m_0=1$ takes the following form in
the no-warping approximation \cite{ourPRB}:
\begin{eqnarray}
\label{lutt}
\vspace{1 cm}
H_L= \left[ \matrix{H_{11} & {\overline \gamma}\sqrt{3}(eB/c)a^2 &
\gamma_3\sqrt{6eB/c}\ k_z a & 0 \cr
\ & H_{22} & 0 & -\gamma_3\sqrt{6eB/c}\ k_z a \cr
\ & \ & H_{33} & {\overline \gamma}\sqrt{3}(eB/c)a^2 \cr
\ & \ & \ & H_{44}} \right],
\end{eqnarray}

\noindent where

\begin{eqnarray}
\nonumber
H_{11}=-(\gamma_1/2-\gamma_2)k_z^2-(eB/c)\left[(\gamma_1+\gamma_2)
\left(a^+ a + \frac{1}{2} \right)+\frac{3}{2}\kappa \right]+V_h(z)+V(x,y), \\
\nonumber
H_{22}=-(\gamma_1/2+\gamma_2)k_z^2-(eB/c)\left[(\gamma_1-\gamma_2)
\left(a^+ a + \frac{1}{2} \right)-\frac{1}{2}\kappa \right]+V_h(z)+V(x,y), \\
\nonumber
H_{33}=-(\gamma_1/2+\gamma_2)k_z^2-(eB/c)\left[(\gamma_1-\gamma_2)
\left(a^+ a + \frac{1}{2} \right)+\frac{1}{2}\kappa \right]+V_h(z)+V(x,y), \\
\nonumber
H_{44}=-(\gamma_1/2-\gamma_2)k_z^2-(eB/c)\left[(\gamma_1+\gamma_2)
\left(a^+ a + \frac{1}{2} \right)-\frac{3}{2}\kappa \right]+V_h(z)+V(x,y).
\end{eqnarray}

\noindent The lower half of matrix (\ref{lutt}) is obtained by
Hermitian conjugation. In Eq.(\ref{lutt}) $a$ is an annihilation operator,
$e$ is a modulus of elementary charge, $\gamma_1$, $\gamma_2$, $\gamma_3$ and
$\kappa$ are the material bulk parameters which are well-known for GaAs.
The hole energy is counted as negative from the upper edge of the valence
band throughout the paper. In the effective mass approximation the $k_z$
component of quasimomentum in (\ref{lutt}) is replaced by its operator form
$k_z=-i\partial / \partial z$. This substitution is performed at $B=0$ and
$V(x,y)=0$ which yields an infinite set of doubly degenerate heavy and light
hole subband energies and eigenfunctions $c_j^{\nu}(z),\nu =1,2,\ldots$.
The $z$-dependent envelope functions $C_j^{\nu}(z)$ at finite $B$ can be
constructed as superpositions of zero-field functions $c_j^{\nu}$
\cite{Broido,Volkov}. Now let the periodic potential (\ref{vxy}) be applied,
corresponding to the rational magnetic flux through the elementary cell with
the area $S=a^2$:
\begin{equation}
\label{pq}
\frac{BS}{\Phi_0}=\frac{BS}{2\pi \hbar c / |e|}=\frac{p}{q}.
\end{equation}

\noindent If the condition (\ref{pq}) is satisfied, any of four components
$\psi^j$ of the vector of hole envelope functions becomes a magnetic Bloch
function classified by $k_x$ and $k_y$ quantum numbers varying in the MBZ
(\ref{MBZ}), and the total hole quantum state can be written as following:
\begin{eqnarray}
\nonumber
\Psi_{k_x,k_y}({\bf r})=
\psi^1_{k_x k_y}({\bf r})\left|\frac32;\frac32\right\rangle+
\psi^2_{k_x k_y}({\bf r})\left|\frac32;-\frac12\right\rangle+ \\
\label{psihole}
+\psi^3_{k_x k_y}({\bf r})\left|\frac32;\frac12\right\rangle+
\psi^4_{k_x k_y}({\bf r})\left|\frac32;-\frac32\right\rangle.
\end{eqnarray}

\noindent The translational properties of each component of the envelope
function (\ref{psihole}) in $(xy)$ plane are the same as for the
single-component electron wavefunction \cite{ourPRB}. In particular, every
component of (\ref{psihole}) satisfies to the Peierls condition (\ref{pei}).
Hence, one can write the components $\psi^j_{k_x k_y}({\bf r})$ of
(\ref{psihole}) as a superposition of the Landau quantum states
\cite{Thouless,Silb,Geisel,DP,ourPRB}, namely
\begin{eqnarray}
\nonumber
\psi^j_{k_x k_y}({\bf r})=\frac{1}{La\sqrt{q}}
\sum_{\nu}C_j^{\nu}(z)\sum_{N}\sum_{n=1}^p d_{j \nu N n}(k_x,k_y)
\sum_{l=-L/2}^{L/2}u_{N}\left(\frac{x-x_0-lqa-nqa/p}{\ell_H} \right)\times
\\
\label{psiho}
\times \exp\left(ik_x\left[lqa+\frac{nqa}{p}\right]\right)
\exp\left(2\pi iy\frac{lp+n}{a}\right)\exp(ik_y y).
\end{eqnarray}

\noindent It should be mentioned that the set of basis functions for the hole
states in magnetic subbands splitted from interacting hole Landau levels has
more complicated structure than those for electrons (the latter is discussed,
for example, in Refs. \cite{Geisel,DP}). Namely, at the absence of periodic
potential the four-component eigenvector of Luttinger Hamiltonian in a single
subband of size quantization $\nu$ has the form \cite{Broido}:
\begin{equation}
\label{fn}
F_{Nk_y}^{\nu}=e^{ik_y}\left(C_1^{\nu}(z)u_{N-2}, \ C_2^{\nu}(z)u_{N},
\ C_3^{\nu}(z)u_{N-1}, \ C_4^{\nu}(z)u_{N+1} \right).
\end{equation}

\noindent In Eq.(\ref{fn}) $u_N(x)$ is a harmonic oscillator wavefunction
which vanish for negative values of its index. Below we shall discuss in
details the structure of expression (\ref{psiho}).

First, we should restrict ourselves to some limited number of size
quantization subbands to be taken into account. In heterojunctions with
typical hole concentration $n_h=5 \times 10^{11} cm^{-2}$ and depletion-layer
density $N_{dep}=10^{15} cm^{-3}$ only the lowest hole subband of size
quantization is occupied \cite{Broido,Volkov}. Hence, in the expression
(\ref{psiho}) for the hole state it seems to be relevant to consider only
several subbands of size quantization neighbouring to the lowest one. Besides,
for each subband of size quantization in Eq.(\ref{psiho}) we take into account
only several Landau levels $N$. During the investigation of hole states
(\ref{psiho}) in this Sec. we consider the first three subbands of size
quantization which corresponds to two heavy- and one light-hole levels.
The basis for the hole state (\ref{psiho}) at $V(x,y)=0$ consists of
the following four-component vectors:
\begin{eqnarray}
\nonumber
e^{ik_y}\left(0,\ 0,\ 0,\ C_4^1(z)u_0 \right), \qquad
e^{ik_y}\left(0,\ C_2^1(z)u_0,\ 0,\ C_4^1(z)u_1 \right), \\
\label{holebas}
e^{ik_y}\left(C_1^1(z)u_0,\ C_2^1(z)u_2,\ C_3^1(z)u_1,
\ C_4^1(z)u_3 \right), \qquad
e^{ik_y}\left(0,\ 0,\ 0,\ C_4^2(z)u_0 \right), \\
\nonumber
e^{ik_y}\left(0,\ 0,\ 0,\ C_4^2(z)u_1 \right),\qquad
e^{ik_y}\left(C_1^2(z)u_0,\ 0,\ 0,\ C_4^2(z)u_3 \right)
\end{eqnarray}

\noindent where the upper index $\nu=1,2$ labels the first or the second
subband of size quantization, and the light-hole components for the second
subband have been removed. It is easy to see that each term in (\ref{holebas})
has the form of (\ref{fn}) with particular values of $\nu$ and $N$. It should
be noted that the group of neighbouring hole levels may not be classified by
a monotonous sequence $N=-1,0,1, \ldots$ in Eq.(\ref{fn}) which is the
fundamental difference between hole and electron Landau level (the latter are
labeled by increasing index). In the presence of periodic potential $V(x,y)$
for which the condition (\ref{pq}) is satisfied, each of Landau levels is
splitted into $p$ subbands. The summation over $n=1,\ldots ,p$ in
Eq.(\ref{psiho}) corresponds to this splitting which is a general feature both
for the electron and hole magnetic Bloch states
\cite{Thouless,Silb,Geisel,DP,ourPRB}. To define the limits for indices
$(\nu, N)$ in (\ref{psiho}), one should fix the $|J;m_J\rangle$ projection
$j=1,2,3,4$ and than take the sum of {\it j}-th components of all vectors in
(\ref{holebas}) with coefficients $d_{j \nu N n}(k_x,k_y)$. The total number
of non-zero components in the set of basis vectors (\ref{holebas}) is equal to
$11$ which is smaller than the total number of avaliable components
$4\times 6=24$ due to the vanish of those components of (\ref{fn}) which have
negative indices. Hence, after substituting the total hole wavefunction
(\ref{psihole}) - (\ref{psiho}) into the Schr\"odinger equation with the
Hamiltonian (\ref{lutt}) one obtains the $11p \times 11p$ eigenvalue problem
for the $11p$ coefficients $d_{j \nu N n}(k_x,k_y)$ in every of $11p$ hole
magnetic subbands.

In the Introduction we've mentioned that the wavefunction of a Bloch electron
has at least $p$ zeros per magnetic cell if the magnetic flux is equal to
$p/q$ of flux quanta, which is a consequence of the Peierls condition
(\ref{pei}). It is interesting to generalize this result for a multi-component
wavefunction. Namely, if $\theta^j_{\bf k}$ denotes the phase of the $j$-th
periodic part
$u_{{\bf k}}^j({\bf r})=\exp(-i{\bf kr})\psi_{{\bf k}}^j({\bf r})$
of the hole component $\psi_{{\bf k}}^j({\bf r})$ defined by (\ref{psiho}),
one can introduce the vorticity $\Gamma_j$ for each component as following:
\begin{equation}
\label{vor}
\Gamma_j=\frac{1}{2\pi}\oint d{\bf l}
\frac{\partial \theta^j_{\bf k}(x,y)}{\partial {\bf l}}
\end{equation}

\noindent where the integration path is taken along the boundary of magnetic
unit cell in the counterclockwise direction. It was mentioned above that the
condition (\ref{pei}) is valid for every component of the vector of hole
envelope functions. So, it is not surprising that the vorticity (\ref{vor})
is equal for all hole components:
\begin{eqnarray}
\label{vor1}
\Gamma_j=-p, \qquad j=1,2,3,4.
\end{eqnarray}

\noindent However, the position inside magnetic
cell and the total number of zeros can be different for each of the
$|J;m_J\rangle$ components due to their particular form of basic functions
and coefficients in (\ref{psiho}). It should be mentioned also that the total
number of zeros per magnetic cell can be greater than $p$ because of the
opposite signs of some vorticities. All of these results are shown on
Fig.1 where the probability distributions of all four hole envelope functions
are plotted at $k_x=k_y=0$ in a non-overlapped magnetic subband at
$p/q=5$.
The zeros are shown as black circles of different size corresponding to their
order (see the inset). One can see that different hole components have the
different number and (in general) the different order of zeros. Some of the
zeros are located on the sides and in the corners of magnetic cell which is
reflected by semi- and quarter-circle areas.
It should be mentioned that despite of the different positions and the number
of zeros for each of $|J;m_J\rangle$ components, the total vorticity per one
magnetic cell (\ref{vor1}) is equal for all components at any $(k_x,k_y)$ in
all magnetic subbands. This result reflects the topological nature of
the wavefunction vorticity (\ref{vor}).
\begin{figure}[t]
\leavevmode
\centering{\epsfbox{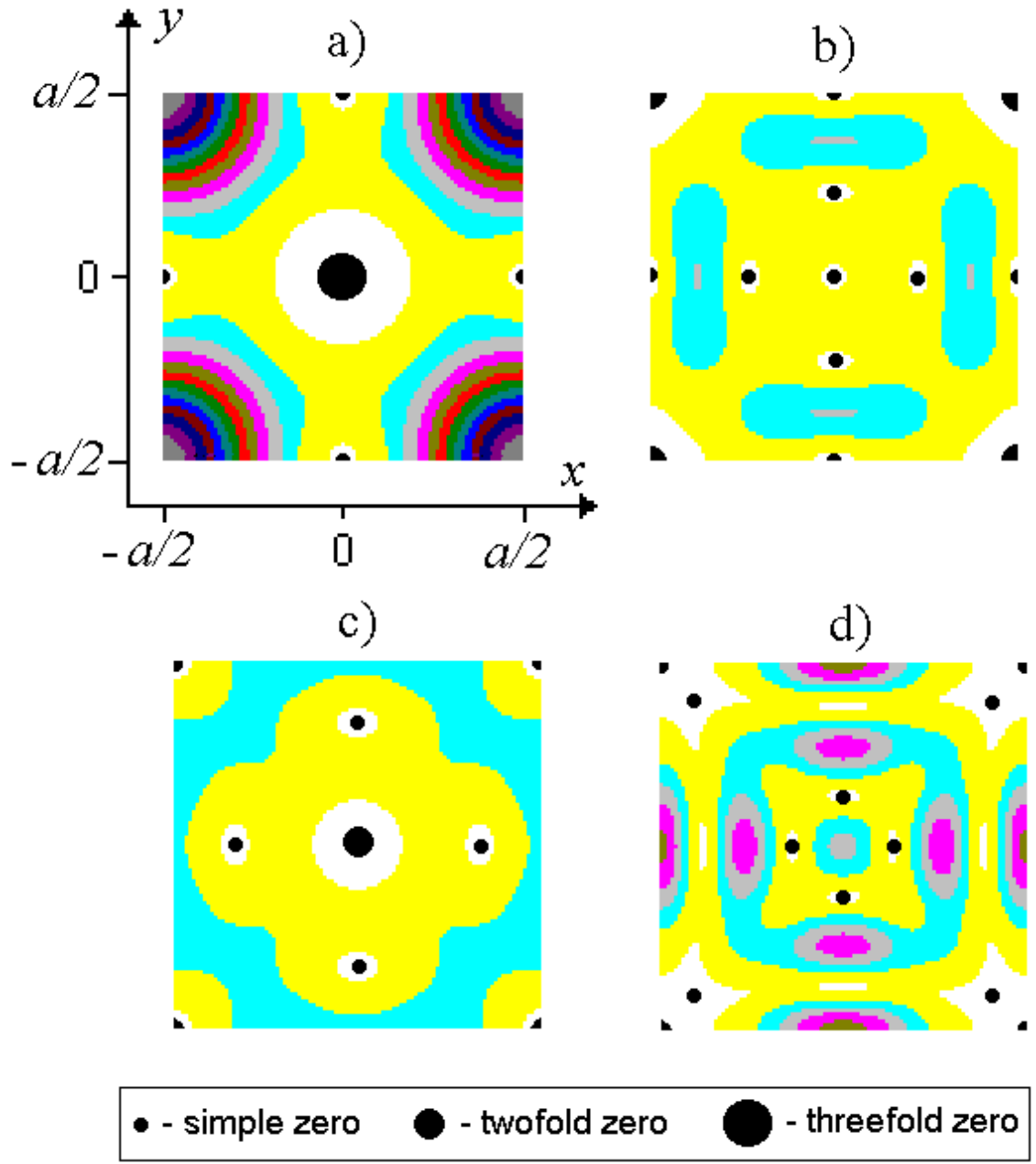}}
\caption{Probability distributions for the $|J;m_J\rangle$ hole components
$J=1 - 4$ (a - d) in a magnetic subband are shown at $p/q=5$ and
$k_x=k_y=0$. Darker areas correspond to the greater values of the
wavefunction modulus. The positions of wavefunction zeros are marked as black
circles with diameter proportional to their order. The zeros located on the
sides and in the corners of magnetic cell which are plotted by semi- and
quarter-circle areas, respectively.}
\label{fqh1}
\end{figure}
\clearpage

\section{Quantization of Hall conductance}

The Hall conductance $\sigma_H$ is quantized in units of $e^2/h$ as soon as
the Fermi energy lays in the energy gap. The value of $\sigma_H$ is determined
by the sum of partial conductances of filled magnetic subbands. Thus, we shall
study at first the Hall conductance of one fully occupied magnetic subband
$\alpha$. In the absence of disorder and at zero temperature, its contribution
to Hall conductance is given by \cite{Thouless,PG,Kohmoto,Usov}
\begin{eqnarray}
\label{sigini}
\sigma_H^{\alpha}=\frac{e^2}{{\pi}^2 \hbar}
\int Im \left\langle \frac{\partial u_{\bf k}}{\partial k_y}
\left| \frac{\partial u_{\bf k}}{\partial k_x} \right. \right\rangle d^2k
\end{eqnarray}

\noindent where $u_{\bf k}=\Psi_{k_xk_y}({\bf r})e^{-i{\bf kr}}$ is the
periodic part of the Bloch function in the current subband $\alpha$. For the
four-component hole state (\ref{psihole}) one obtaines from (\ref{sigini})
the following expression for $\sigma_H$:
\begin{eqnarray}
\label{sighole}
\sigma_H^{\alpha}=\frac{e^2}{{\pi}^2 \hbar}
\sum_{j=1}^4\int Im \left\langle \frac{\partial u^j_{\bf k}}{\partial k_y}
\left| \frac{\partial u^j_{\bf k}}{\partial k_x} \right. \right\rangle d^2k
\end{eqnarray}

\noindent where $u^j_{\bf k}=\Psi^{(j)}_{k_xk_y}({\bf r})e^{-i{\bf kr}}$
and $\Psi^{(j)}_{k_xk_y}({\bf r})$ is defined by (\ref{psiho}). In this
Section we shall focus on the magnetic subbands originating from the lowest
subband of size quantization and thus will ignore the second heavy-hole
subband, omitting the last three vectors in (\ref{holebas}). This will reduce
the number of non-zero components from $11$ to $7$. After substituting
$u^j_{\bf k}$ into (\ref{sigini}) and taking into account the orthogonality
and normalization of the basis functions in (\ref{psiho}), one may express the Hall conductance
(\ref{sighole}) through the partial derivations of the components
$d_{j \nu N n}(k_x,k_y)$ desribing the quantum state. For brevity, in
the following we shall replace the set of indices $(j \nu N n)$ by
a single index $n=1,\ldots, 7p$ which runs sequentially all the required
values. We get
\begin{eqnarray}
\label{sigdn}
\sigma_H=\frac{e^2}{{\pi}^2 \hbar}
\int Im\left[\frac{i}{2}\ell_H^2+\sum_{n=1}^{7p}
\frac{\partial d_n^*}{\partial k_y}\frac{\partial d_n}{\partial k_x}
\right]d^2k
\end{eqnarray}

\noindent The expression (\ref{sigdn}) has been initially derived \cite{Usov}
for the Hall conductance of a magnetic subband splitted from a single electron
Landau level. We claim that (\ref{sigdn}) is valid for the case of several
interacting electron or hole levels as long as the spectrum is non-degenerate.
The differences with the single-level problem is only in the size of the
matrix equation which is now equal to $N\cdot p$ instead of $p$.
The orthogonality and normalization of the basis functions used for
construction of the hole envelope function (\ref{psiho}) is of the same kind
as the basis set for a single-level problem. This feature allows us to expand
directly the approach used by Usov for the case of several interacting levels.
So, we use the expression (\ref{sigdn}) for calculations of the Hall
conductance for the hole Landau levels which are coupled even at zero
superlattice potential by the off-diagonal elements of the Luttinger
Hamiltonian.

It is evident from (\ref{sigdn}) that for calculation of the Hall
conductance one should study first the analytical properties of coefficients
$d_n$ as the functions of quasimomentum. First, one can transform the 2D
integral (\ref{sigdn}) into a 1D contour integral. In order to simplify the
integration and to reduce it to the summation of the winding numbers over the
singularities (see the right side of (\ref{usov})), one has to introduce
the extended magnetic Brillouin zone (EMBZ) which is derived from the
previously determined magnetic Brillouin zone (\ref{MBZ}) by extending it
$p/q$ times in the $k_y$ direction:
\begin{eqnarray}
\label{ext}
-\pi /qa \le k_x \le \pi /qa, \quad  -p\pi /qa  \le k_y \le p\pi /qa.
\end{eqnarray}

\noindent It was shown by Usov \cite{Usov} that the integration along the
"boundaries" of the EMBZ (\ref{ext}) gives no impact to the value of
$\sigma_H$ which is explicitly determined only by the contour integrals
around the singularities (winding numbers).
We shall briefly repeat the outline of the derivation of this result.
One can choose a representaion for which one of the components of the vector
${\bf d}=(d_1,\ldots d_{Np})$, say, $d_1$, is real. The singular points
${\bf k}_m$, $m=1,2\ldots$ are thus determined as points where
$d_1({\bf k}_m)=0$. The other components will be written as
$d_j=|d_j|\exp(i\theta_j)$, $j=2,3,\ldots$. The winding numbers appear to
be equal for any of these components. To be specific, let us consider the
calculation of the winding numbers for $d_2$.
At a particular singular point ${\bf k}_m$ its winding number $S_m$ can be
calculated as an algebraic sum of rotations (modulo $2\pi$) of the vector with
the components $(Re d_2, \em Im d_2)$. A typical behavior of $d_1$ and both
$Re d_2$ and $Im d_2$ in two magnetic subbands is illustrated on Figures 2-3
for the magnetic flux $p/q=3/2$, $a = 80$ nm  and the amplitude of periodic
potential $V_0 = 0.7$ meV which corresponds to the case of non-overlapped
magnetic subbands.
\begin{figure}[t]
\leavevmode
\centering{\epsfbox{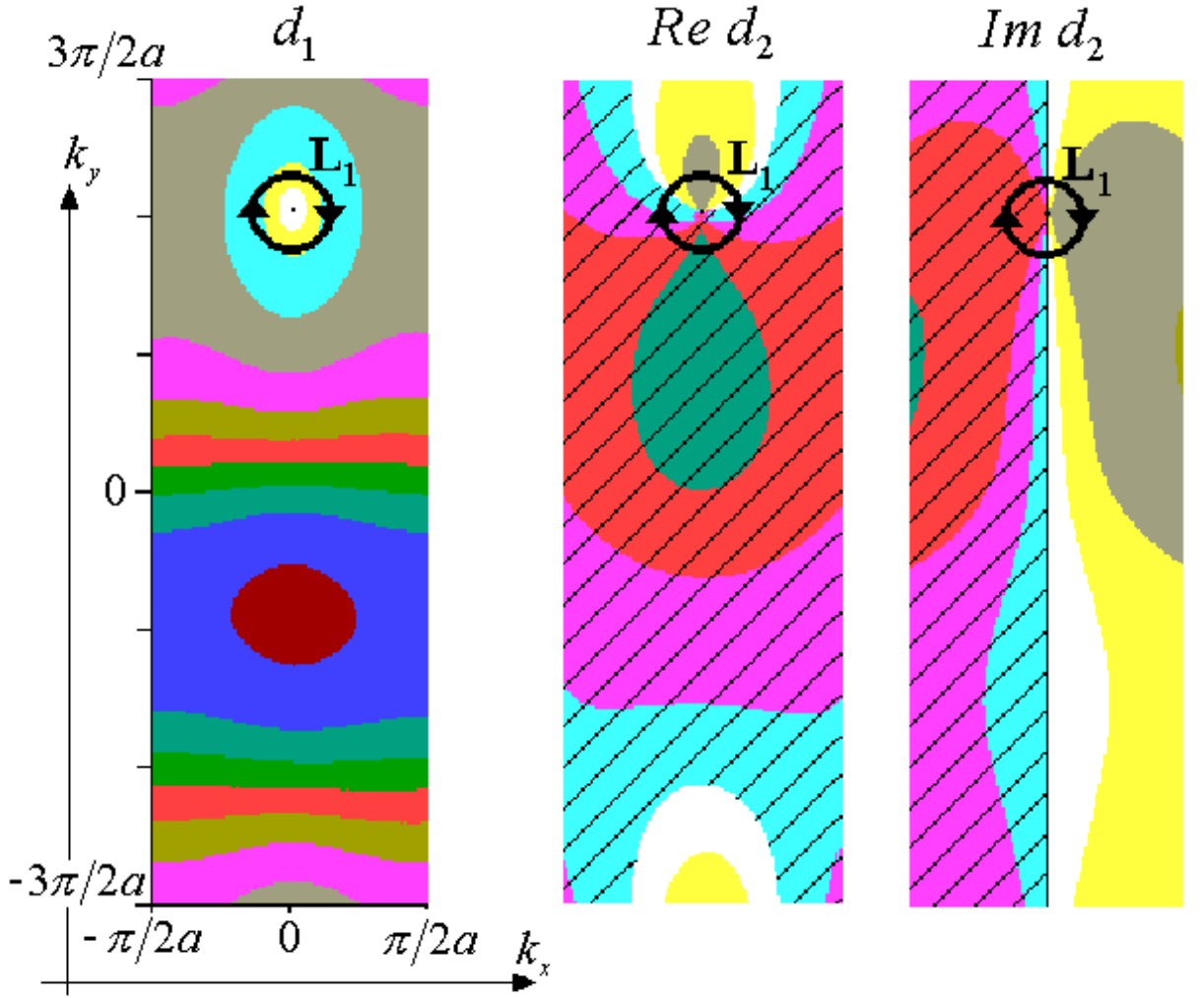}}
\caption{Typical behavior of two components of the eigenvector
${\bf d}$ describing the hole quantum state in the representation with real
$d_1$. The eigenvector is taken for a magnetic subband with the Hall
conductivity equal to $+1$ in units $e^2/h$. The magnetic flux $p/q=3/2$,
$a = 80$ nm  and the amplitude of periodic potential $V_0 = 0.7$ meV which
corresponds to the case of non-overlapped magnetic subbands. Darker areas
correspond to the greater values of the $d_{1,2}$ modulus, and the negative
parts are shaded with lines. The contour $L$ and the direction of circulation
around the singularity are shown schematically.}
\label{fqh2a}
\end{figure}

The contours $L_{1,2}$ of circulations around the
singularities are shown schematically. It is evident from Figures 2-3 that
while approaching the singular point where $d_1 = 0$  which is marked by black
dot, both real and imaginary parts of $d_2$ have different limits depending on
the direction in $(k_x,k_y)$ plane and thus do not have a true limit in this
point. The impact of the component $d_j$ at a singular point ${\bf k}_m$ to
the Hall conductance is proportional to $|d_j|^2S^j_m$ where $S^j_m$
is the winding number for the component $d_j$. It was shown \cite{Usov} that
for all components $j=2,3,\ldots, Np$ the winding numbers are equal,
$S^j_m=S_m$. So, the summation over all components gives the impact to
the Hall conductance provided by a singular point ${\bf k}_m$:
\begin{equation}
\nonumber
\sum_j|d_j|^2S^j_m=S_m\sum_j|d_j|^2=S_m,
\end{equation}

\noindent where we've used the normalization of the vector
${\bf d}=(d_1,\ldots d_{Np})$. As soon as the winding numbers
are calculated, the Hall conductance of a particular magnetic subband is given
by Eq.(\ref{usov}). By examining the expression (\ref{usov}), one can mention
that the first term in the square brackets is just the contribution of
one out of $p$ subbands to the free-electron Hall conductivity. The remaining
term in (\ref{usov}) is stipulated by the presence of periodic potential
and by a non-trivial topology of the wavefunction in the EMBZ (\ref{ext}).
At least one singular point can be found for every magnetic subband when
${\bf k}$ runs over EMBZ. As a result, the expression in brackets in
(\ref{usov}) is always an integer, which is the first Chern class of magnetic
subband \cite{Kohmoto}.
\begin{figure}[t]
\leavevmode
\centering{\epsfbox{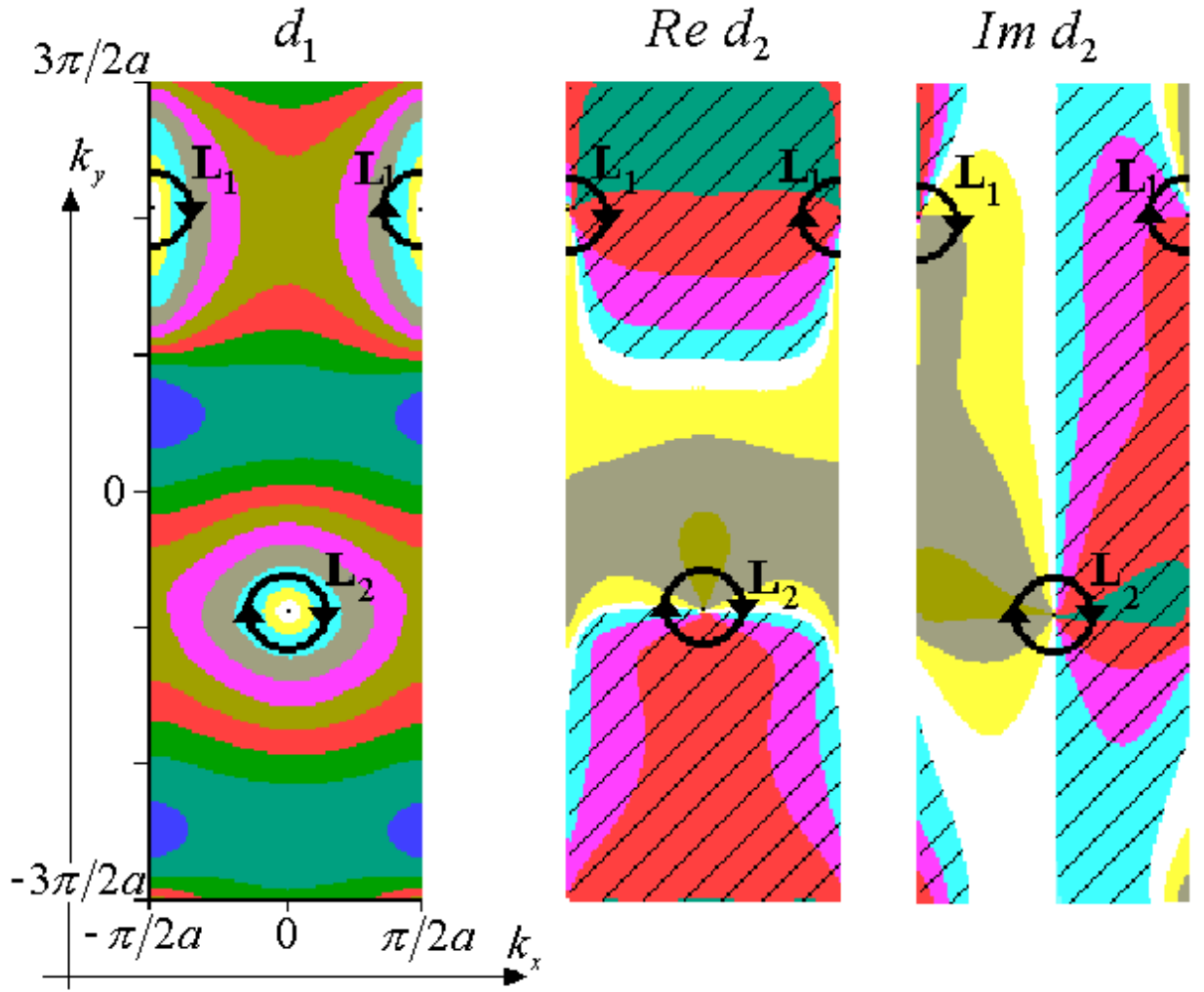}}
\caption{Same as Fig.3 but for a subband with Hall conductivity equal
to $-1$.}
\label{fqh2b}
\end{figure}

The quantization of $\sigma_H$ as a function of
the number of filled magnetic subbands (or, equivalently, of the position of
the Fermi level) is shown on Figures 4 - 5 both for non-overlapped and
overlapped magnetic subbands.
When the amplitude $V_0$ of the periodic potential (\ref{vxy}) is smaller
than the distance $\Delta E_{12}$ between neighbouring Landau levels, none
of the subbands are overlapped (see Fig.4) and possible deviations in the
quantization of $\sigma_H$ from the sequences obtained for the Hofstadter
"butterfly" \cite{Thouless} are caused by the non-equidistant character of
hole Landau levels which leads to hole subband spectra with another structure
than the "butterfly" for electrons \cite{ourPRB}. The quantization of
$\sigma_H$ in the gaps between non-overlapped subbands as a function of Fermi
energy is plotted on Fig.4. It should be noted that when the Fermi level is
swept through a subband centered at $E_n$ (or through a region of overlapped
subbands, see below), the Hall conductivity interpolates smoothly between the
adjacent quantized values. These values are shown by solid lines on Figures
4 - 5 and the interpolation curves are the dashed lines.
\begin{figure}[t]
\leavevmode
\centering{\epsfbox{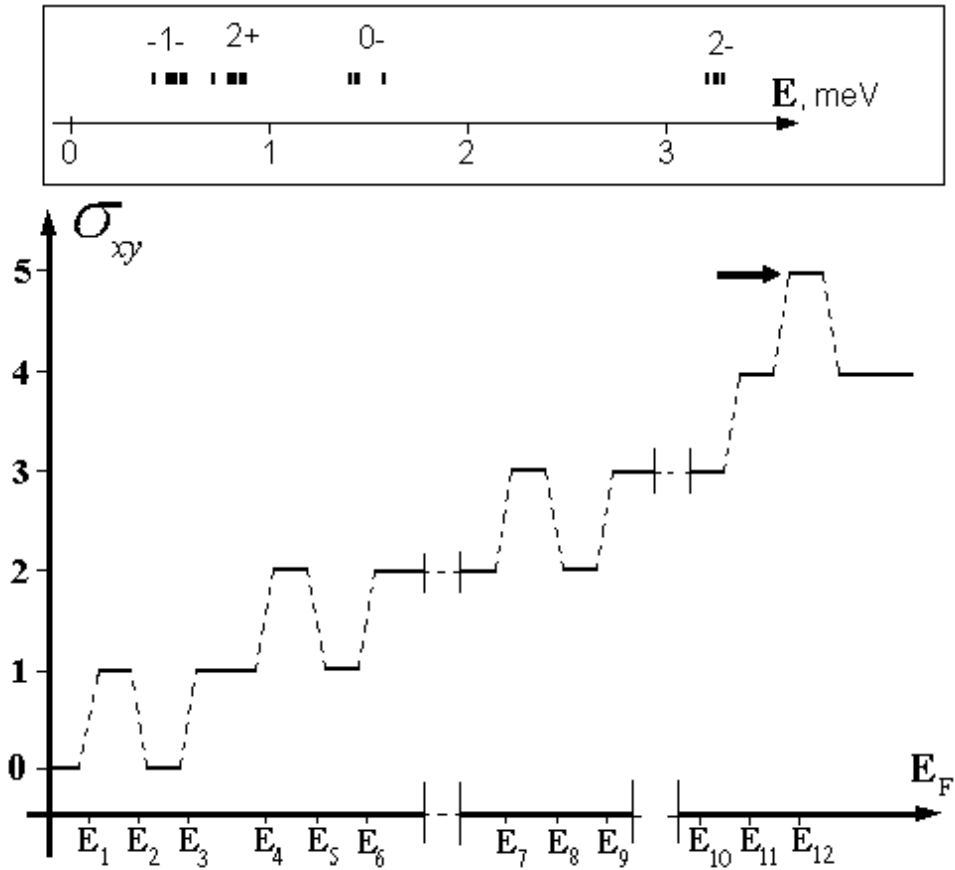}}
\caption{(Top inset) Non-overlapped hole magnetic subbands originating
from four hole Landau levels with indices $N=-1,0,2$ and dominating spin
projections $\pm$. The magnetic flux $p/q=3/2$ and the amplitude of periodic
potential $V_0=0.7$ meV. (Bottom) Quantization of $\sigma_H$ as a function of
the Fermi level position. The energies $E_i$ of the centeres of subbands on
the inset are shown chematically, and the dashed lines serve as a guide to the
eye. The arrow indicate the deviation from the quantization sequence for
Hofstadter "butterfly".}
\label{fqh3}
\end{figure}

If the amplitude $V_0$ is increased, the neighbouring magnetic subbands from
different hole Landau levels can touch each other at some point in the MBZ.
This touch means that the degeneracy of the spectrum has occured, and the
application of the expression (\ref{usov}) is invalid. However, the further
increase of $V_0$ leads to the repulsion of the touched subbands and to the
breaking of the degeneracy. So, one can use (\ref{usov}) at higher $V_0$
when the condition $V_0>\Delta E$ is satisfied and some of magnetic subbands
are overlapped but the spectrum remains to be non-degenerate. The spectrum for
$V_0=3$ meV is shown on the inset of Fig.5. One can see that the number and
the maximum width of gaps on Fig.5 has decreased with respect to the system
of non-overlapped subbands on Fig.4 which will reduce the number and maximum
width of Hall plateaus. For convenience, on Fig.5 we've labeled the remaining
gaps and the corresponding Hall plateaus by numbers. Again the dashed line
serves as a guide to the eye and it qualitatively shows the impact to the Hall
conductance provided by those subbands which are fully occupied at the current
position of $E_F$ when it is swept through the region of overlapped subbands.
The unusially high and low, even negative values of $\sigma_H$ are marked by
arrows. It should be stressed that the differences between the behavior of
$\sigma_H$ on Figures 4 and 5 are provided by only two changes in the
$\sigma_H^{\alpha}$ for subbands $\alpha=4$ and $\alpha=8$ (see Fig.4).
We found that these subbands have been degenerated at some intermediate values
of $V_0$ which are greater than on Fig.4 but lower than on Fig.5. According to
the topological point of view \cite{Kohmoto,Novikov}, the subband touches have
caused the exchange of the Chern classes $\Delta c=\pm q$ between these
subbands where $q=2$ in our examples. It can be easily seen that such exchange
($-2$ for subband 4 and $+2$ for subband 8) exactly transforms the
quantization shown on Fig.4 to the dependence on Fig.5. We suppose that the
novell features of the quantization of Hall conductance in laterally modulated
hole gas which have been discussed in this paper can be detected in transport
experiments.
\begin{figure}[t]
\leavevmode
\centering{\epsfbox{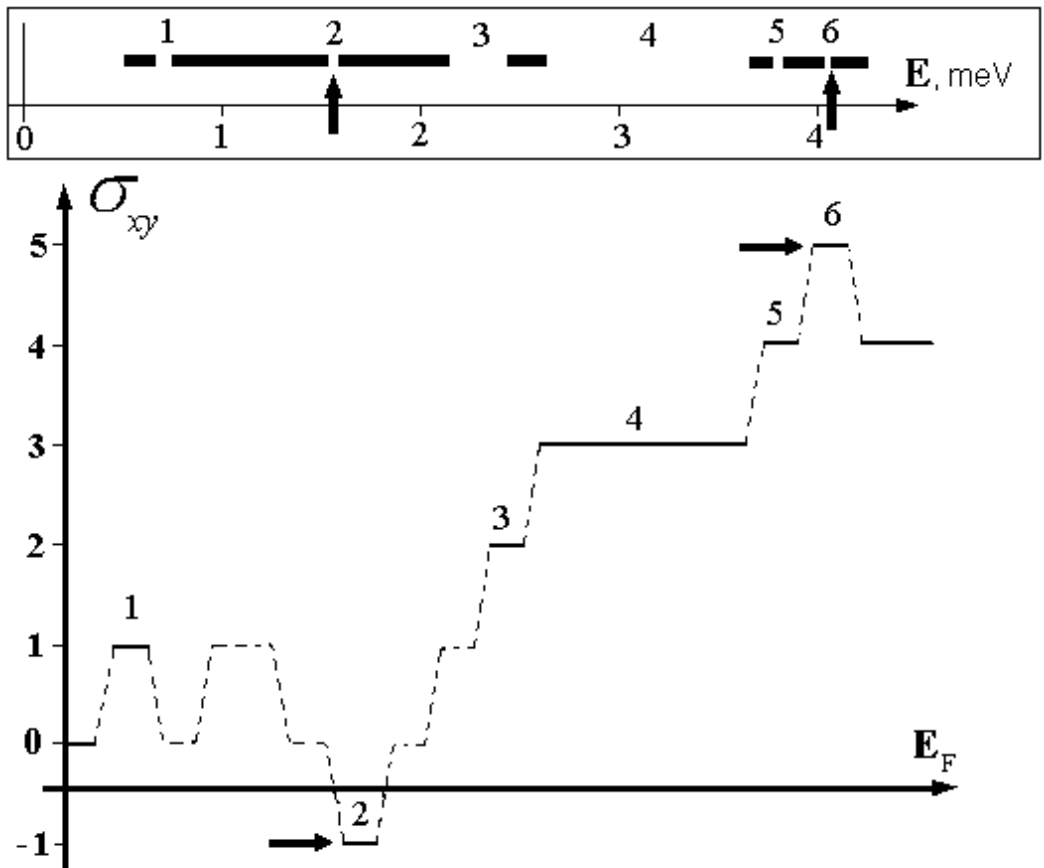}}
\caption{(Top inset) Overlapped hole magnetic subbands originating from
the same hole Landau levels as on Fig.~\ref{fqh3} but splitted by stronger
periodic potential with $V_0=3$ meV. The numbers label energy gaps. (Bottom)
Quantization of $\sigma_H$ in the energy gaps labeled on the inset as
a function of the Fermi level position. The deviations from the quantization
sequence for Hofstadter "butterfly" are marked by arrows.}
\label{fqh4}
\end{figure}

\section{Summary}

We presented a new method of calculation of the Hall conductance in a 2D hole
gas affected by lateral periodic potential which is a generalization of
the method derived by Usov \cite{Usov} for a system where the charged particle
is described by a four-component eigenfunction of the Luttinger Hamiltonian.
An unusual behavior of the Hall conductance in comparence to the well-known
dependance for Hofstadter "butterfly"as a function of the Fermi energy
is found for holes. The explanation of this effect is proposed by evaluating
the role of the off-diagonal terms of the Luttinger Hamiltonian which provide
a highly non-equdistant character of hole Landau levels. The quantization of
the Hall conductance is investigated both at weak and at strong periodic
potential. In the latter case the magnetic subband mixing is taken into
account which leads to the exchange of the Chern numbers between magnetic
subands, changing their impact to the Hall conductance. The experimental
observation of the differences between the quantization of Hall conductance
in n- and p-type heterojunctions is discussed.

\section*{Acknowledgments}

We thank D. Weiss, R.R. Gerhardts and D. Pfannkuche for useful discussions and
M. Kohmoto for sending us the hardcopy of his paper \cite{Kohmoto}. This work
was supported by the RFBR (Grants No. 01-02-17102, 02-02-06440), by the
Russian Ministry of Education (Grants No. E00-3.1-413, UR 0101.020) and by the
BRHE Program (Project REC - 001).

\end{document}